\documentclass{article}

\usepackage[final,nonatbib]{neurips_2019}
\usepackage{graphicx}
\graphicspath{ {./NeuRIPS2019/} }

\usepackage[utf8]{inputenc} 
\usepackage[T1]{fontenc}    
\usepackage{hyperref}       
\usepackage{url}            
\usepackage{booktabs}       
\usepackage{amsfonts}       
\usepackage{nicefrac}       
\usepackage{microtype}      

\title{ABOUT ML: Annotation and Benchmarking on Understanding and Transparency of Machine Learning Lifecycles}

\author{%
  Deborah I. Raji \footnotemark \\
  Partnership on AI\\
  San Francisco, CA \\
  \texttt{deb@partnershiponai.org} \\
  \And
  Jingying Yang \footnotemark[\value{footnote}] \\
  Partnership on AI \\
San Francisco, CA \\
  \texttt{jingying@partnershiponai.org} \\
}

\begin{document}

\maketitle
 \footnotetext{Equal contribution.}
\begin{abstract}
We present the "Annotation and Benchmarking on Understanding and Transparency of Machine Learning Lifecycles" (ABOUT ML) project as an initiative to operationalize ML transparency and work towards a standard ML documentation practice. We make the case for the project's relevance and effectiveness in consolidating disparate efforts across a variety of stakeholders, as well as bringing in the perspectives of currently missing voices that will be valuable in shaping future conversations. We  describe the details of the initiative and the gaps we hope this project will help address.
\end{abstract}

\section{Introduction}

When AI is deployed within human systems, complications can arise, often leading to unanticipated and undesirable  consequences \cite{selbst2019fairness}. At times, the low compatibility of an AI system to the human context arises from a simple lack of communication. If certain details are not made explicit between system developers and impacted stakeholders, then the system can be unknowingly misused and its results become misinterpreted, untrustworthy and difficult to hold accountable \cite{thomaz2006transparency, veale2017logics}. 

It is for this reason that transparency is emerging as a major priority for organizations around the world \cite{jobin2019artificial}. Laws like the Fair Credit Reporting Act (FCRA), the Equal Credit Opportunity Act (ECOA), and the General Data Protection Regulation (GDPR), government procurement processes like Canada’s Algorithmic Impact Assessment, and engineering practice at IBM \cite{hind2018increasing}, Microsoft \cite{gebru2018datasheets} and Google \cite{mitchell2019model} in particular highlight the shift towards documentation-based approaches to transparency as a practical mechanism to achieving more trustworthy machine learning deployments. Information-based approaches involve a recorded exchange between the system developer and users in order to reach a shared understanding of the known details of the system and its intended function \cite{veale2017logics}. As a far more accessible, inexpensive and simple solution to ethical AI deployment challenges than the often more complex and abstract fairness interventions available, transparency through documentation is a promising practical intervention that can integrate into existing workflows to provide clarity in decision making for users, external auditors, procurement departments, and other stakeholders alike. 

In this paper, we go over the details of a practical transparency approach to making ML systems more human-compatible. We discuss the evidence of the acknowledged importance of transparency as a value to industry and government stakeholders and then summarize the details of the approach of transparency through documentation. We then present the "Annotation and Benchmarking on Understanding and Transparency of Machine Learning Lifecycles" (ABOUT ML) project as a future resource and standard to operationalize this principle and consolidate efforts across stakeholders. Finally, we acknowledge the ongoing challenges as we move forward with the project.

\section{Demand for Transparency in ML Systems}

In machine learning, models can encode complex and unintuitive relationships between inputs and outputs, making it challenging for human operators to naturally infer the details of what guided the process leading to a particular outcome. As a result, many organizations include transparency as a core value in their AI principle statements. Of 50 AI principle statements documented through the Linking AI Principles (LAIP) project, 94\% (47) explicitly mention transparency\cite{zeng2018linking}. Similarly, 87\% and 88\% of principle statements surveyed in two other concurrent studies reference “transparency” \cite{jobin2019artificial}. In fact, transparency is often highlighted as the most frequently occurring principle in these survey studies, and has been named "the most prevalent principle in the current literature” \cite{jobin2019artificial}. 

That being said, the intricacy and difficulty of translating the high-level ethical ideal of transparency into concrete engineering processes and requirements has been repeatedly referenced as a major challenge. Although study after study confirms that meaningful progress cannot be made until ethical ideals are operationalized \cite{whittlestone2019role, mittelstadt2019ai, greene2019better}, the inconsistency with which high level principles such as transparency are interpreted across different contexts, organizations and even teams makes it difficult to design consistent practical interventions. This lack of practical theory serves as a roadblock to facilitating outside auditing from interested parties looking to hold AI system developers accountable, and can impede or slow down the responsible deployment of these models \cite{greene2019better, selbst2019fairness}.

\section{Transparency Through Documentation}
One simple and accessible approach to increasing transparency in ML lifecycles is through an improvement in both internal and external documentation norms and processes. This is in fact among the most common of reported fairness interventions implemented in industry and government \cite{holstein2019improving}. For an increasingly concerned public or auditing organization, thorough, externally distributed documentation on ML systems is essential to earning and maintaining trust, and minimizing the misuse of these systems. External documentation and reporting standards can also assist practitioners in making the case within their organizations to allocate the necessary resources to more thoroughly incorporate ethics into their AI projects. Internal documentation is also vital, serving to improve communication between collaborating teams. Internal documents build employee trust by outlining the nature of an individual or team’s contribution to an overall system, giving opportunity for ethical objections and a more meaningful understanding of the impact of their personal participation in the creation of an end product. Beyond the artifact, however, the process of documentation itself is inherently valuable, as it prompts critical thinking about the ethical implications at every step in the ML lifecycle and encourages adherence to the set of steps required to understand and report a complete picture of system capabilities, limitations, and risks.


Documentation for transparency is thus both an artifact (in this case, a document with details about the ML system, similar to a nutrition label on food) and a process (in this case, a series of steps people follow  in order to create the document). Both of these interpretations are at the core of the initial effort of the ABOUT ML initiative, which focuses on developing documentation to clarify the details of specific ML systems, for the sake of improving the transparency of that system.

\section{Research Themes on Documentation for Transparency}

In order to define the characteristics and intended uses of the system, there are well-researched sets of documentation questions already available, through various disparate research efforts. These past research efforts differ greatly from one another and are often specific to a particular domain such as Natural Language Processing \cite{bender2018data}, or geared towards a specific element of the system, such as the dataset \cite{gebru2018datasheets}. 
As these documentation templates are often modeled on those used in other industries, such as safety data sheets from the electronics industry \cite{gebru2018datasheets} or nutrition labels from the food industry \cite{holland2018dataset}, the suggested templates vary widely in length and appearance, ranging from a single concise page of succinct statements \cite{mitchell2019model} or a set of symbols and visualizations \cite{holland2018dataset, kelley2009nutrition} to upwards of 10 pages of detailed prose and graphs \cite{gebru2018datasheets}. Whether the documentation is meant for internal or external consumption also impacts length and contents, as internal documentation can be more detailed and thus longer. There is thus currently great variability in the past research attempts to inform the documentation questions and format in machine learning development, and a need for consolidated guidance for the community with respect to documentation best practices.

Despite these differences, there certain themes from past work to pay attention to. A common focus across templates, even outside of data-specific work, is on clarifying the details of data provenance, for both the training and testing data used in ML development \cite{gebru2018datasheets, bender2018data, mitchell2019model}. Documentation questions across papers consistently address the risks that arise at various stages of data creation and distribution, with the goal of encouraging practitioners to reflect on ethical concerns at every stage including data use and release, and some templates placing additional focus on specific risks like privacy \cite{kelley2009nutrition}. Another recurring theme in the related work is on the importance of clarifying the model's intended use and objectives. A major stated goal for these templates is to allow the team to articulate initial objectives, so they can refer back to these goals to ensure ongoing consistency with their declared intentions. Model- and system-level documentation efforts that emerged from earlier work on data documentation, in particular introduce questions more specific to the definition of overall operational objectives and design decisions within a broader system \cite{mitchell2019model, hind2018increasing}. Some work has gone further to suggest a legal advantage to declaring intended use cases and ethical concerns, as it can provide grounds for legally restricting third party misuse \cite{benjamin2019towards}.

Also, even with a broad range of contributors to the current body of work, certain voices are consistently missing. For instance, the perspective of civil rights organizations and government representatives is often not included in this work, despite their very specific expectations for transparent systems and the important role of documentation in system auditing. There is also no record of what those impacted by the deployed ML systems would like to see reported in documentation. Getting that feedback requires formalizing the inclusion of the perspectives of those most affected by the ML system, especially people from traditionally marginalized and underrepresented communities. As most corporate perspectives have been produced by large multi-national technology companies, it is also important to diversify our understanding of what industry-appropriate documentation practice looks like, and how the process should accommodate less resourced engineering workflows.  


\section{ABOUT ML}

The ABOUT ML project\footnote{More information on the project available at: https://www.partnershiponai.org/about-ml/} is an iterative, multistakeholder process to collaboratively create best practices via input from diverse perspectives, with the goal of translating those best practices into industry norms. Although just completing its primary stages, the ABOUT ML project is a promising model for standardizing the operationalization of other common AI ethics principles, and prioritizing inclusion throughout the process. By catalyzing cross-organizational collaboration, the goal is to translate ethical ideals such as transparency into research-backed tools for a variety of stakeholders.

Organizations have already begun to implement documentation recommendations from research publications, and such work is beginning to influence documentation requirements in regulation and engineering practice \cite{holstein2019improving}. However, there is no consensus on which practices work best and still a lack of understanding of which basic information needs to be disclosed in an ML system. In fact, the definition of transparency itself is highly contextual. As a result, there is currently no standardized process for the documentation of machine learning systems. Each team that wants to apply the research summarized above to improve transparency in their ML systems via documentation must address the entire suite of questions about what transparency means for their team, product, and organization given their specific goals and constraints, with little formal guidance. 
 
The goal of ABOUT ML is thus to consolidate past efforts and condense that work into meaningful guidelines and templates to support documentation practice in machine learning. The process is modeled after iterative ongoing processes to design internet standards (such as W3C, IETF, and WHATWG) and includes a public forum for discussion and a place to submit any proposed changes. The resulting template recommendations can serve as a head start to those looking to implement these strategies. Rather than a rigid list of requirements, ABOUT ML will offer a summary of recommendations and practices that is mindful of the variance in transparency expectations, in order to guide teams to identify and address their context-specific challenges. 

This project also serves as a method to accelerate academic progress on the topic by pooling insights more quickly, sharing resources, and reducing the redundancy of efforts. The success and quality of the eventual ABOUT ML output depends on engagement and buy-in from a wide range of relevant stakeholders. The hosting organization is committed to investing in the resources necessary to seek out other groups undertaking transparency initiatives and incorporate their lessons into ABOUT ML recommendations.

With ABOUT ML, we aim to create a platform for teams and individuals to discuss and share experiences alongside researchers, civil society organizations, advocacy groups, users, and other people impacted by AI technology by creating and maintaining an online forum and a concurrent public comment process. In order to make this process as inclusive and robust as possible, we have designed a standardization process with two key design elements: a Steering Committee and the Diverse Voices methodology.

Keeping up with the latest developments in research and practice, we recruited ~30 experts, researchers and practitioners from a diverse set of partner organizations to serve on the ABOUT ML Steering Committee. This committee will guide the updating of ABOUT ML drafts based on submitted public comments, new research developments in the field and advances in reported practices. They will approve new releases by “rough consensus,” which is commonly used by other multi-stakeholder working groups \cite{resnick2014consensus}. The steering committee is representative of organizations from a broad set of perspectives, including civil society organizations, non-profits, large and small corporations as well as academic institutions.

To ensure that diverse perspectives — especially those from communities historically excluded from technology decision-making — contribute to any ABOUT ML recommendations, we are engaging with the Tech Policy Lab at the University of Washington to conduct Diverse Voices panels for the ABOUT ML project \cite{young2019toward}. This methodology was designed to gather feedback from stakeholders who are impacted by a technology policy but whose perspectives might not otherwise be consulted in its formation. Thus, for each iteration of the ABOUT ML template, this Diverse Voices panel feedback will inform the final edits incorporated before release, providing an alternate perspective to the Steering Committee input.

\section{Current Ongoing Challenges \& Gaps}

When attempting to implement the recommended documentation guidelines, a number of common challenges arise. For instance, a deeper study of current institutional structures and missing enablers of transparency interventions would serve as an excellent foundation for future pilot and implementation phases of ABOUT ML. Additional gaps for further work include curating a consolidated set of documentation questions, defining a shared understanding of what is necessary to consider an ML system transparent, and agreeing on an equitable process to empower more stakeholders to have input on what goes into documentation. There are also inherent known limitations to transparency as a mechanism for trust \cite{ananny2018seeing}, as well as information security, intellectual property and customization concerns to accommodate when designing for safe system disclosures.  Despite these challenges, the ABOUT ML project is an important first step in bringing together disparate and at times even economically competitive groups together to move the industry towards more transparent and compatible ML systems.

\section{Conclusion}
For ML systems to preserve privacy, ensure fairness, and reduce bias they must first be developed and deployed within a framework that provides accountability. The industry attention on the AI principle of transparency provides an opportunity to finally take action towards creating more human-compatible ML systems. The ABOUT ML project hopes to channel this enthusiasm into practical results by lowering the barrier to integrating documentation processes into any team and workflow.

Although decentralized experimentation has begun on documentation as a transparency intervention, there lacks adequate guidance to implement or make sense of the available recommendations, particularly across a diversity of contexts and interests. ABOUT ML can hopefully evolve into a framework to offer proper support of the attempted real-world implementations of this approach, and thus support the increased transparency of deployed ML systems overall. 

\section{Appendix}
Below is an overview of the ABOUT ML process to standardize ML documentation for transparency. 

The project is currently on schedule. Release version 0 is in the process of Steering Committee and Diverse Voices, after undergoing a public comment period. Release version 1 is expected at the beginning of 2020, and will include more a more formal set of recommendations.

More details can be found at the following web address: https://www.partnershiponai.org/about-ml/ 

We encourage anyone interested in participating in the project to get in touch for futher updates.

\begin{figure}[h]
  \centering
  \includegraphics[width=\textwidth]{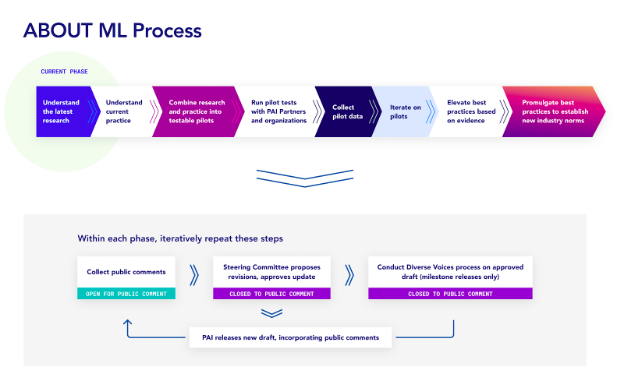}
  \caption{Overview of ABOUT ML project's lifecycle.}
  
  \centering
  \includegraphics[width=\textwidth]{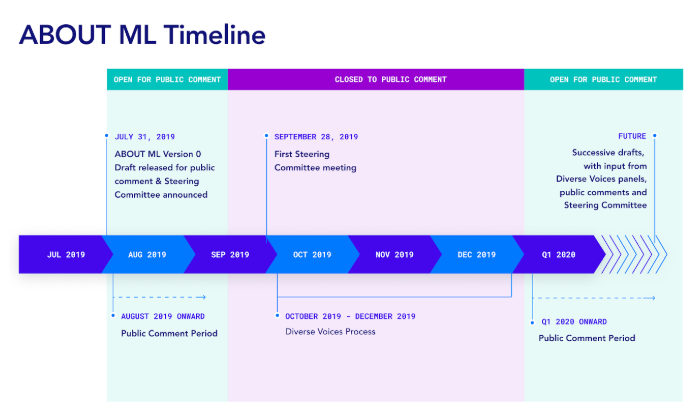}
  \caption{Current timeline for ABOUT ML.}
\end{figure}

	\bibliographystyle{plain}
	\bibliography{neurips_2019}

\end{document}